\documentclass[letterpaper, 10 pt, conference]{ieeeconf}  

\IEEEoverridecommandlockouts                              

\usepackage{graphics} 
\usepackage{graphicx}
\usepackage{epsfig} 
\usepackage{mathptmx} 
\usepackage{times} 
\usepackage{amsmath} 
\usepackage{amssymb}  
\usepackage{comment}
\usepackage{xcolor}
\usepackage{placeins}
\usepackage{float}
\usepackage{cite}
\usepackage{multirow}
\usepackage{xcolor}
\usepackage{gensymb} 
\usepackage{fancyhdr}
\usepackage{hyperref}
\usepackage{textcomp}
\usepackage[letterpaper,left=0.75in, right=0.75in, headheight=10pt, headsep=10mm]{geometry}
\title{\LARGE \bf
Fast Marching based Tissue Adaptive Delay Estimation for Aberration Corrected Delay and Sum Beamforming in Ultrasound Imaging}

\author{Asif M S$^{1*}$, Gayathri Malamal$^{1*}$, Madhavanunni A N$^{1}$, Vikram Melapudi$^{2}$, Rahul V$^{2}$, \\ Abhijit Patil$^{2}$, Rajesh Langoju$^{2}$ and Mahesh Raveendranatha Panicker$^{1}$
 \thanks{*Equal Contributions}
\thanks{$^{1}$Asif M S, Gayathri Malamal, Madhavanunni A N and Mahesh Raveendranatha Panicker (Email:mahesh@iitpkd.ac.in) are with the Center for Computational Imaging and Department of Electrical Engineering, Indian Institute of Technology Palakkad.}       
\thanks{$^{2}$Vikram Melapudi, Rahul V, Abhijit Patil and Rajesh Langoju are with GE Healthcare Bangalore.}}

\begin{document}
\begin{titlepage}
    \vspace*{\fill}
\fontsize{15}{18}\selectfont\textcopyright { 2023 IEEE. Personal use of this material is permitted. Permission from IEEE must be obtained for all other uses, in any current or future media, including reprinting/republishing this material for advertising or promotional purposes, creating new collective works, for resale or redistribution to servers or lists, or reuse of any copyrighted component of this work in other works.}
    \vspace*{\fill}
\end{titlepage}
\maketitle
\pagestyle{fancy}
\thispagestyle{fancy}
\fancyhead{}
\renewcommand{\headrulewidth}{0pt}
\fancyhead[L]{\fontsize{10}{10} \selectfont \textcolor{black}{This article is accepted in the $45^{th}$ Annual International Conference of the IEEE Engineering in Medicine and Biology Society
(EMBC 2023)}}
\fancyfoot{}

\begin{abstract}
Conventional ultrasound (US) imaging employs the delay and sum (DAS) receive beamforming with dynamic receive focus for image reconstruction due to its simplicity and robustness. However, the DAS beamforming follows a geometrical method of delay estimation with a spatially constant speed-of-sound (SoS) of 1540 m/s  throughout the medium irrespective of the tissue in-homogeneity. This approximation leads to errors in delay estimations that accumulate with depth and degrades the resolution, contrast and overall accuracy of the US image. In this work, we propose a fast marching based DAS for focused transmissions which leverages the approximate SoS map to estimate the refraction corrected propagation delays for each pixel in the medium. The proposed approach is validated qualitatively and quantitatively for imaging depths of upto $\sim$ 11 cm through simulations, where fat layer induced aberration is employed to alter the SoS in the medium. To the best of authors' knowledge, this is the first work considering the effect of SoS on image quality for deeper imaging. 
\newline
\indent \textit{Clinical relevance}— The proposed approach when employed with an approximate SoS estimation technique can aid in overcoming the fat induced signal aberrations and thereby in the accurate imaging of various pathologies of liver and abdomen. 
\end{abstract}

\section{INTRODUCTION}
In a typical ultrasound (US) system, the images are reconstructed through a delay and sum (DAS) beamformer, which delays the signals to compensate for the round trip propagation in the medium, apodized, and summed across the transducers to form the final output. The estimation of the round trip delay requires knowledge of the speed of sound (SoS) in the medium. The human tissues are heterogenous and are known to have SoS variations of up to 10\%. Irrespective of this, the conventional DAS uses a spatially constant SoS of $\sim$1540 m/s for delay compensation, leading to imprecise travel time estimation and thus degrading the resolution, contrast and overall accuracy of the US image \cite{Szabo2004DiagnosticEdition}.   

Improvements in resolution and contrast of US images are seen in the literature, where estimation of propagation delay based on refraction correction are used \cite{Wang2013TranscranialStudy, Ali2022DistributedEstimates}. However, they either take into account the spatial variations of SoS in certain regions of the medium, such as the skull \cite{Wang2013TranscranialStudy}, or they are specific to certain methods of transmission, such as synthetic aperture  \cite{Ali2022DistributedEstimates}. Moreover, these methods are only demonstrated for depths of upto 5 cm, which is not the case for imaging tissues at a larger depth as in liver or abdomen.

In this work, a fast marching (FM) based novel delay estimation for DAS receive beamforming with focused transmissions is proposed and studied through simulations with heterogenous SoS distributions for imaging depths of upto 11 cm. The qualitative and quantitative results demonstrate the capability of FM based DAS through noticeable improvements in the image quality by minimizing tissue distortions even at greater imaging depths when compared to the conventional DAS based on geometrical delay estimation. 
\section{METHODS}
\subsection{Review of Transmit and Receive Beamforming}
\begin{figure}[!t]
\centering
\includegraphics[width=0.6\columnwidth]{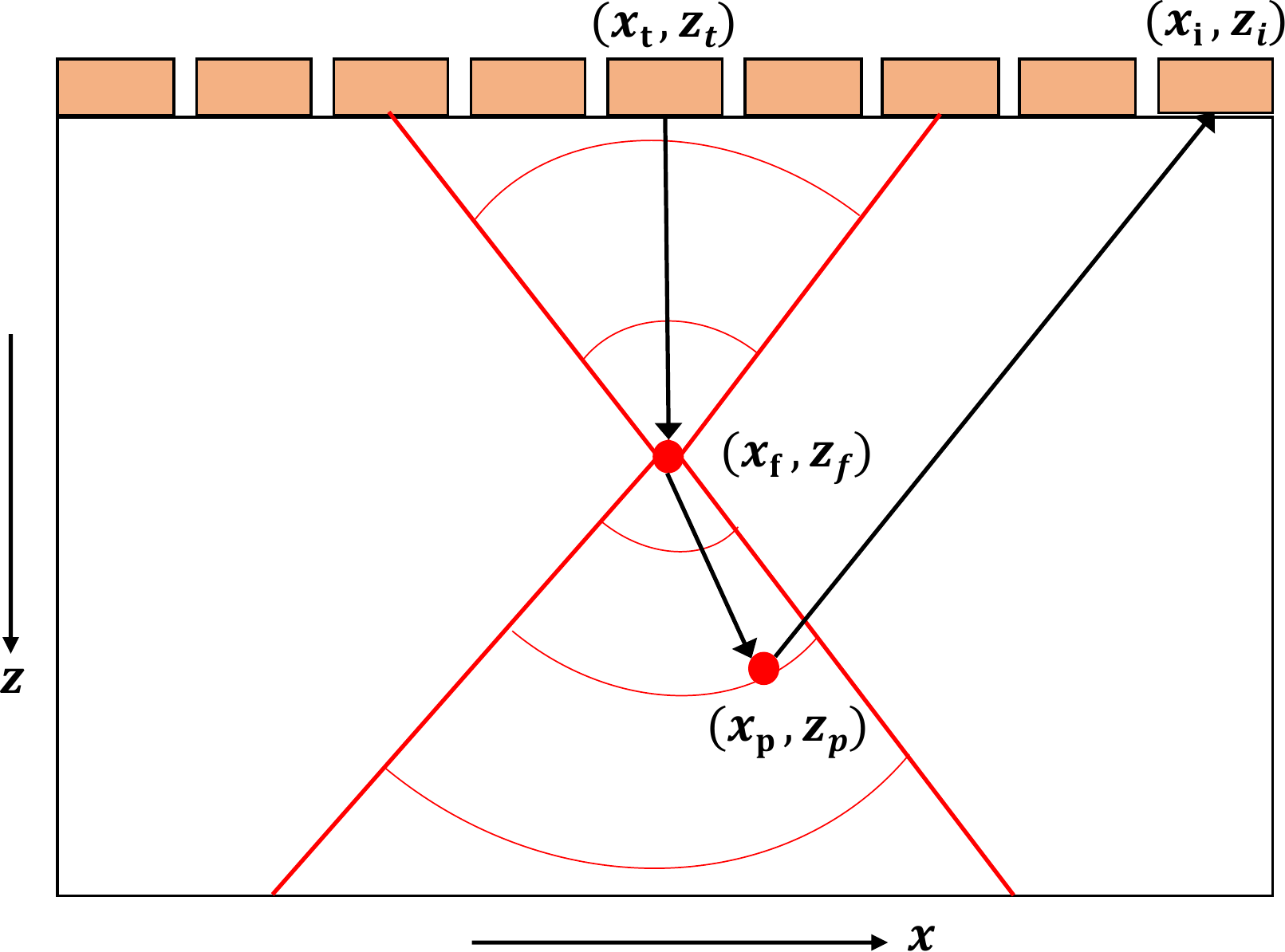}
\caption{Geometry for delay calculation for focused transmission, with center of the transmit aperture at $(x_{t},z_{t})$ and focal point at $(x_{f},z_{f})$. The receiving element is denoted by $(x_{i},z_{i})$. }
\label{fig:virtual_source_geometry}
\end{figure}

Assume that a uniform linear array transducer with $N_c$ elements, with a total array width $L$ and transmit wavelength ($\lambda$) placed at depth ($z = 0$) is used for imaging. For focused transmit beamforming, a sub-aperture of elements in the transducer array are electronically delayed appropriately to converge the US beam to a prefixed point/depth called focus. This creates a single line of image across the depth for every transmit, and multiple transmissions with varying focal points are used to reconstruct the whole image \cite{Szabo2004DiagnosticEdition}. An illustration of focused transmit beamforming is shown in Fig. \ref{fig:virtual_source_geometry} where a sub-aperture of the transducer array centered at  $(x_{t},z_{t})$ is excited to focus the US beam at $(x_{f},z_{f})$. However, the contribution of each focused transmission to reconstructing every pixel $(x_{p},z_{p})$ in the medium might alternatively be estimated in place of the standard single scan line method and is adopted in this work \cite{HoelRindal2018AModel}. 

The transmitted US signals interact with the medium, and reflections are generated due to acoustic impedance mismatches at various interfaces of the medium, which are detected by the same transducer array. To reconstruct the images from the echo signals, conventional US imaging systems employ the delay and sum (DAS) receive beamforming with dynamic receive focusing. The backscattered signals arriving at the transducer array are compensated for the time of flight of propagation in the medium, apodized (weighted) and coherently summed across the transducer elements and compounded across the transmissions to obtain the beamformed signal corresponding to every pixel in the medium \cite{HoelRindal2018AModel}. For a set of $M$ focused transmissions, this is mathematically expressed in \eqref{eq:DAS}, 
\begin{equation}
   \label{eq:DAS}
    y_{DAS}(p) = \sum_{j=1}^{M} \sum_{i=1}^{N_{c}} W_{i}(p) s_{i}(\tau_{p}(i,j))
\end{equation}
where, $y_{DAS}$ is the beamformed signal for the pixel $p$, $W_i$ is the apodization weight of the $i^{th}$ transducer element and $s_{i}$ is the signal received in the $i^{th}$ transducer element compensated with the round trip propagation delay $\tau_{p}$ which is the sum of the transmit $\tau_{tx}$ and the receive $\tau_{rx}$ delays in the medium as given in \eqref{eq:foc_total_delay},
\begin{equation}
    \tau_{p} (i,j) = \tau_{tx}(j,p) + \tau_{rx} (i,p)
    \label{eq:foc_total_delay}
\end{equation}
The transmit delay $\tau_{tx}$ is the sum of the travel times from the transmit center $(x_{t},z_{t})$ to the focal point $(x_{f},z_{f})$ and from the focal point (or the virtual source) $(x_{f},z_{f})$ to the pixel of interest $(x_{p},z_{p})$ \cite{HoelRindal2018AModel} which is expressed as in \eqref{eq:foc_tx_total_delay}- \eqref{eq:foc_tx_pixel_delay}.
\begin{equation}
    \tau_{tx} (j,p) = \tau_{foc}(j) + \tau_{foc\_p} (p)
    \label{eq:foc_tx_total_delay}
\end{equation}
where,
\begin{equation}
    \tau_{foc} (j)= \frac{\sqrt{\left(x_t-x_f\right)^2+\left(z_t-z_f\right)^2}}{c}
    \label{eq:foc_tx_delay}
\end{equation}

\begin{equation}
     \tau_{foc\_p}(p) = (-1)^{\left(z_p<z_f\right)} \frac{\sqrt{\left(x_p-x_f\right)^2+\left(z_p-z_f\right)^2}}{c}
     \label{eq:foc_tx_pixel_delay}
 \end{equation}  
 
The receive delay is however, completely independent of the transmit delay and is a factor of the Euclidean distance of the pixel from each receiving transducer element $(x_{i},z_{i})$ which is calculated as in \eqref{eq:rx_delay},
\begin{equation}
   \tau_{rx}(i, p) = \frac{1}{c} \sqrt{(x_{i}-x_{p})^2 + (z_{i}-z_{p})^2}
   \label{eq:rx_delay}
\end{equation}
In all cases, $\textit{c}$ is the spatially constant SoS of the medium.
 
\subsection{Proposed Fast Marching (FM) based DAS}
\begin{figure}[!b]
\centering
\includegraphics[width=\columnwidth] {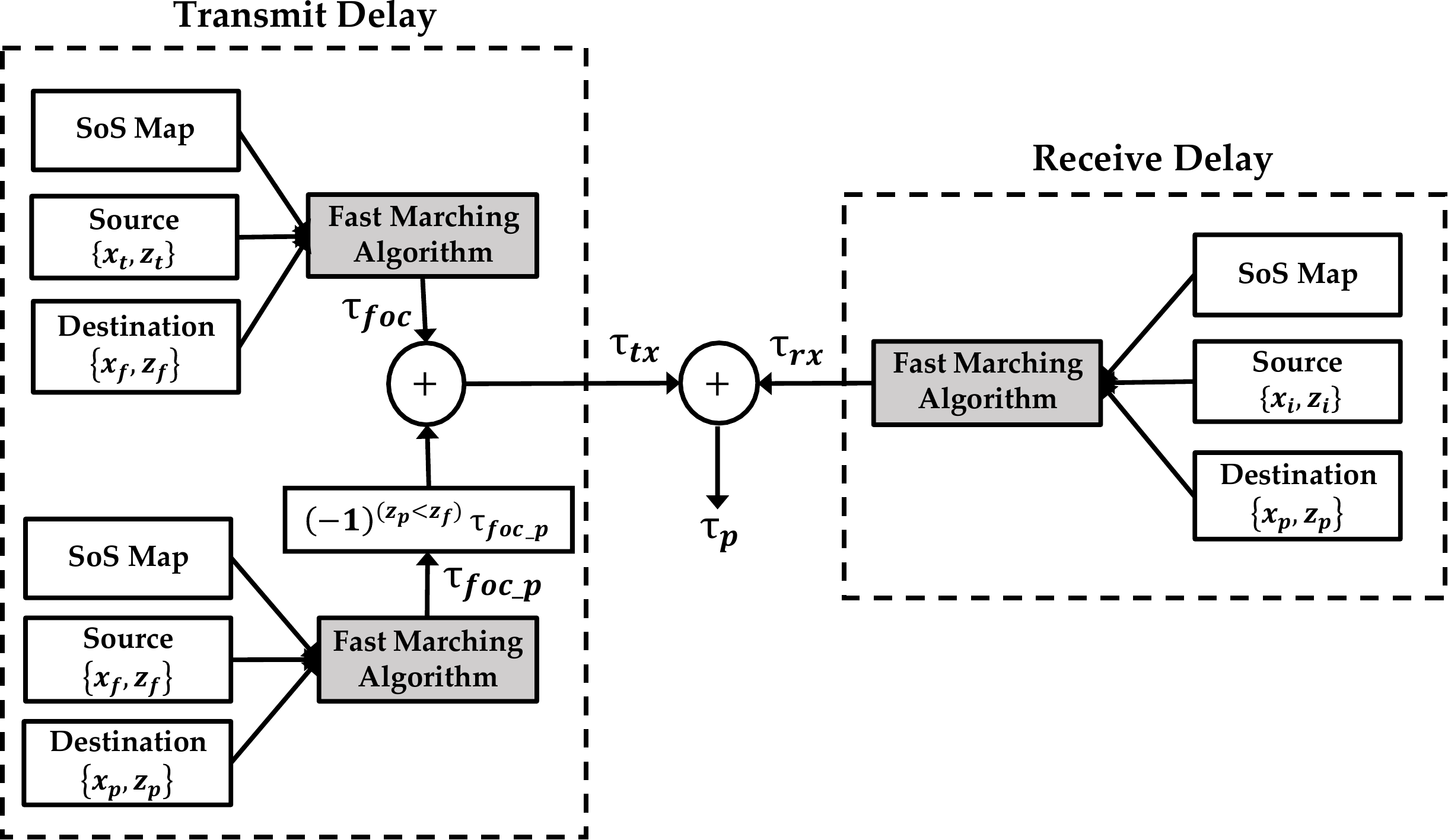}
\caption{The proposed method of delay estimation for DAS receive beamforming for focused transmissions.}
\label{fig:Proposed_Approach_Workflow}
\end{figure}  

The transmit and receive delay estimations  \eqref{eq:foc_tx_total_delay}-\eqref{eq:rx_delay} in conventional DAS \eqref{eq:DAS} assumes a constant SoS $\textit{c}$ irrespective of the spatially varying SoS distribution in the medium. This leads to severe degradation of resolution and affects the consistency of the US image. In this work, we propose a novel fast marching (FM) based technique that accounts for the SoS heterogeneities in the medium for the precise calculation of the travel times for every pixel in the medium for focused transmissions. Consider $c(x_{p},z_{p})$ to be the SoS at the location $(x_{p},z_{p})$ in the medium. The one-way refraction corrected delay ($\tau'$) between $(x_{p},z_{p})$ to any other location in the medium could be solved with the eikonal equation \cite{Wang2013TranscranialStudy, Ali2022DistributedEstimates}, as given by \eqref{eq:eikonal},
\begin{equation}
\label{eq:eikonal}
    \sqrt{\left(\frac{\partial \tau'}{\partial x_p}\right)^2+\left(\frac{\partial \tau'}{\partial z_p}\right)^2}=\frac{1}{c(x_{p},z_{p})}
\end{equation}

The eikonal equation could be accurately solved by using the fast marching (FM) algorithm \cite{Sethian20123-DMethod} which inputs the SoS map of the medium along with the source and the destination locations to output the propagation delay between the points. The workflow for the proposed approach is given in Fig. \ref{fig:Proposed_Approach_Workflow}. The transmit delay \eqref{eq:foc_tx_total_delay} estimation requires two individual boundary conditions to initialize the FM algorithm, as seen from Fig. \ref{fig:Proposed_Approach_Workflow}. For estimating $\tau_{foc}$, where the source is the transmit center $(x_{t},z_{t})$ and the destination is the focal point $(x_{f},z_{f})$, the FM algorithm is initialized with $\tau'(x_{t},z_{t})=0$. For estimating the second component of ($\tau_{tx}$) which is $\tau_{foc\_p}$, the focal point $(x_{f},z_{f})$ is the source and the pixel of interest $(x_{p},z_{p})$ the destination, FM is subjected to the boundary condition $\tau'(x_{f},z_{f})=0$. Further, to obtain the travel time from  $(x_{f},z_{f})$ to $(x_{p},z_{p})$, $\tau_{foc\_p}$ is made negative if $z_{p} < z_{f}$ or retained otherwise as in \eqref{eq:foc_tx_pixel_delay}. The receive delay where the source is the receiving transducer element $(x_{i},z_{i})$ (which is the same as assuming source at $(x_{i},z_{i})$) and the destination is the pixel $(x_{p},z_{p})$ could be computed by solving the eikonal equation subject to the initial condition $\tau'(x_{i},0)=0$. The output of the proposed approach is the sum of the refraction corrected transmit and receive delays, that could be used with the DAS beamforming \eqref{eq:DAS}.  
\begin{figure}[!t]
\centering
\includegraphics[width=0.8\columnwidth]{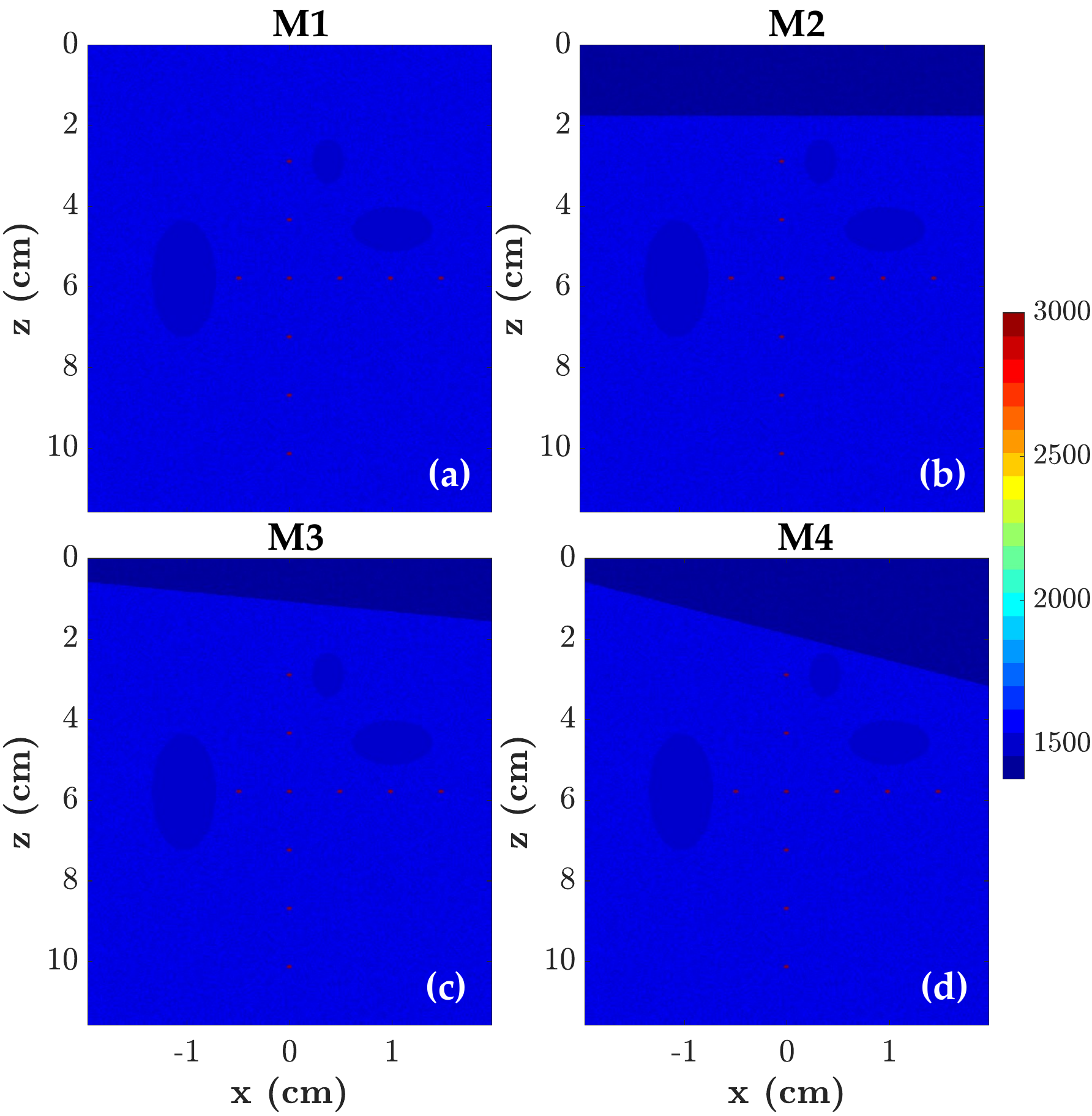}
\caption{The SoS maps of the four scenarios modeled with \textit{k}-Wave US toolbox (a) M1 (no fat layer) (b) M2 (straight fat layer) (c) M3 (fat layer inclined at 10$\degree$) (d) M4 (fat layer inclined at 25$\degree$).}
\label{fig:SoS_Maps}
\end{figure}  
\subsection{Simulation Setup}
Four scenarios (M1, M2, M3, and M4) with heterogenous SoS distributions are simulated with the \textit{k}-Wave US toolbox \cite{Treeby2010K-Wave:Fields} in MATLAB. A 2-D grid ($xz$ plane) of 38.5 mm x 120 mm with a spatial step size of 75 $\mu$m and a 1-D linear array transducer with a pitch of 0.3 mm, a width of 39.4 mm and a center frequency of 3 MHz are modeled. A perfect matching layer of 2 mm is added at the grid boundaries to satisfy the boundary conditions. The transducer array is placed parallel to the $xz$ plane and centered at the origin ($x = z = 0$). The SoS maps for the four cases are shown in Fig. \ref{fig:SoS_Maps}. The mediums consist of three cysts (CY1, CY2, and CY3) of different shapes and orientations located at $(x, z)$ = (-1.3, 5.3), (0.4, 2.6), and (1.3, 4.2) cm respectively and with a constant SoS of 1540 m/s. A set of six point targets with a SoS of 3000 m/s spaced at 1.5 cm are placed axially at $x$ = 0 cm and four laterally at $z$= 5.3 cm. Further, as seen in Fig. \ref{fig:SoS_Maps}, heterogeneities are created with varying orientations of fat mimicking layers, at the superficial depths of the medium. M1 is modeled without the fat layer, a straight fat layer is introduced in M2, an inclined fat layer of 10$\degree$ and 25$\degree$ is modeled in M3 and M4 respectively with an average SoS of 1400m/s. The remaining regions in the mediums contain a random distribution of scatterers (with standard deviation of $1\%$) with an average SoS of 1540 m/s. The raw radio frequency (RF) data is generated with US transmissions focused at approximately half the depth (i.e., 60 mm) for a set of 128 focal points along the lateral direction and a simulation time step of $\sim$ 1.8 ns and a Courant-Friedrichs- Lewy number of 0.05. The transmit apodization used is a hanning window with an $F-number$ of 2.

\subsection{Evaluation Metrics}
\subsubsection{Geometric Distortion Score (GDS)}
The image quality is quantified by estimating the shift in the point targets from their true location. The degree of geometrical correctness is calculated with a metric called geometric distortion score (GDS) which compares each point target's -6dB lateral beamwidth and peak intensity position against its actual location. The point target is considered to be distorted if the maximum distance of the point target from its true position is greater than one wavelength, and assigned a score of 0 and 1 if otherwise \cite{Liebgott2016Plane-WaveUltrasound}.

\subsubsection{Generalized Contrast to Noise Ratio (gCNR)}
The generalized contrast-to-noise ratio (gCNR)  \cite{Rodriguez-Molares2020TheDetectability} is estimated to quantify the contrast by finding the overlap area of the probability density functions inside and outside the cyst regions in the simulated mediums as \eqref{eqn_gCNR}, 
\begin{equation} 
\label{eqn_gCNR}
gCNR = 1- \int  \min \{f_{CY}, f_{BG}\}
\end{equation}
where, $f_{CY}$ and  $f_{BG}$ are respectively the probability density functions of the signals from the cyst and the background regions.

\begin{figure*}[!t]
\centering
\includegraphics[width=0.75\linewidth]{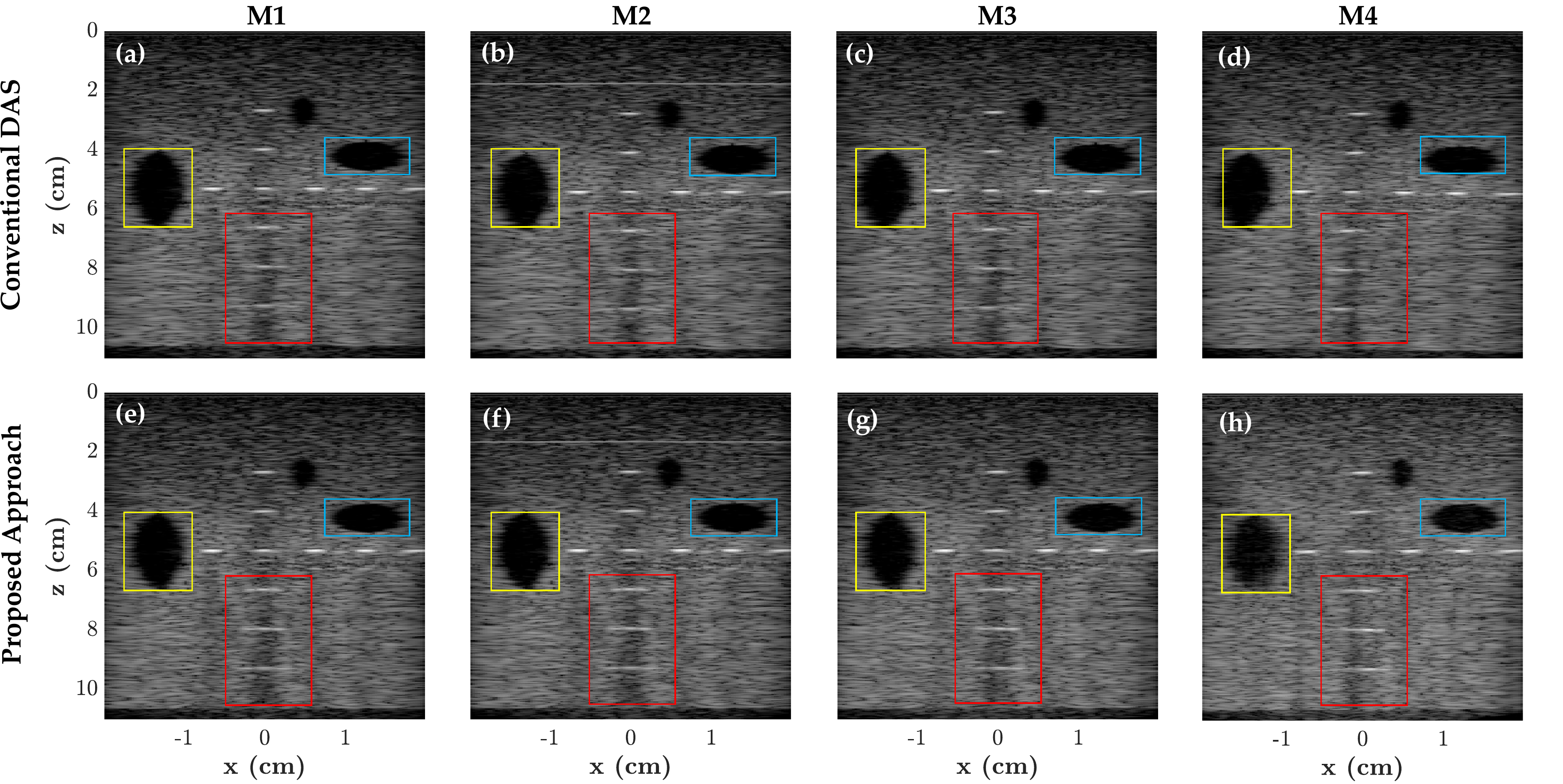}
\caption{Beamformed images with conventional DAS beamforming for (a) M1  b) M2 (c) M3 (d) M4. Beamformed images with the DAS beamforming after delay correction using the proposed approach for (e) M1  (f) M2 (g) M3 and (h) M4. The images are shown for a 60 dB dynamic range.}
\label{fig:BF_Images}
\end{figure*}

\section{RESULTS AND DISCUSSION}
The simulated RF datasets are beamformed with DAS employing the conventional geometrical delay estimation with a spatially constant SoS of 1540 m/s and further with the proposed FM based delay estimation. The envelope detected, log compressed final images are shown in Fig. \ref{fig:BF_Images}.  Fig. \ref{fig:BF_Images} (a)-(d) are the DAS beamformed images with the conventional geometrical delay estimation. It is observed that, as we progress from M1 to M4, the positions of the cysts, particularly CY3 (blue rectangle) and the pins (red rectangle) appear shifted in the beamformed images. The shifts are more prominent in M3 (Fig. \ref{fig:BF_Images}(c)) and M4 (Fig. \ref{fig:BF_Images}(d)), which have inclined fat layers at the superficial depth. Moreover, there is an evident shear in the shape of CY1 in (Fig. \ref{fig:BF_Images}(c) and (d), yellow rectangle). It is also observed that the fat layer interface is clearly distinguishable only in M2 where it is a straight layer and not in M3 and M4 where the layer is inclined. The results are validated quantitatively with GDS and gCNR metrics, which are presented in Table \ref{tab:compare_metric}, row 1. The GDS is computed as the mean of geometric distortion scores over the 10 point targets in the medium, and the gCNR is presented for CY1. In M1, a mean GDS of 0.7 indicates that 7 out of the 10 point targets are correctly localized, which reduces to 0.1 and 0.2 in M2 and M3 respectively. However, in M4, none of the points could be geometrically localized, resulting in a mean GDS of 0. Considering the true location of CY1, circular regions centered at $(x, z)$ = (-1.3, 5.3) cm and $(x, z)$ = (-1.3, 3) cm with a diameter of  1 cm are selected inside CY1 and in the background region respectively for gCNR evaluation. From Table \ref{tab:compare_metric}, it is observed that highest gCNR is observed for the medium M1, which gradually reduces by 15\% (0.67) in the medium M4. This is because gCNR is measured with respect to the true location of CY1, and the shear that occurs with CY1 causes the cyst region to be deviated out of the evaluation region as we progress from M1 to M4. In other words, the observed distortions are due to the errors in the SoS which lead to imprecise delay estimations. This is aggravated in the presence of superficial fat layers, especially when they are inclined as in Fig. \ref{fig:BF_Images} (c) and (d), that may add an extra aberration in the travel times at deeper depths. 
\begin{table}[!t]
\centering
\caption{Comparison of Mean GDS and gCNR between the conventional DAS and FM based DAS}
\vspace{-5pt}
\label{tab:compare_metric}
\begin{tabular}{cccc}
\hline
\hline
\multirow{4}{*}{\textbf{\begin{tabular}[c]{@{}c@{}}Beamforming \\ Method \end{tabular}}} &
\multirow{4}{*}{\textbf{\begin{tabular}[c]{@{}c@{}}Medium \\  \end{tabular}}} &
 \multirow{4}{*}{\textbf{\begin{tabular}[c]{@{}c@{}}Mean \\ GDS \end{tabular}}} &
 \multirow{4}{*}{\textbf{\begin{tabular}[c]{@{}c@{}}gCNR\end{tabular}}} \\
   &    &    \\
   &    &    \\ \hline
\multirow{4}{*}{\textbf{\begin{tabular}[c]{@{}c@{}}Conventional DAS\end{tabular}}}& M1 & 0.7 & 0.79 \\ 
& M2 & 0.1 & 0.77 \\ 
& M3 & 0.2 & 0.75 \\ 
& M4 & 0 & 0.67 \\ \hline
\multirow{4}{*}{\textbf{\begin{tabular}[c]{@{}c@{}}FM based DAS\end{tabular}}}& M1 & 0.7 & 0.76  \\
& M2 & 0.7 & 0.77 \\ 
& M3 & 0.9 & 0.75 \\ 
& M4 & 0.6 & 0.75 \\ \hline
\hline
\end{tabular}
\end{table}

To implement the proposed FM based delay estimation, the ground truth SoS maps obtained from \textit{k}-Wave (Fig. \ref{fig:SoS_Maps}) is smoothed and median filtered to remove sudden changes in speed of sound and provided as input to the FM algoirthm along with the sources and destinations for the transmit and receive delays as in Fig. \ref{fig:Proposed_Approach_Workflow} \cite{AccurateCentral}. The DAS images for the four datasets, using the proposed FM based delay estimation, are shown in Fig. \ref{fig:BF_Images} (e)-(h). It is seen that the proposed FM based DAS is able to correct the distortions in the cysts and the pins. The shift in the pins and the CY3 are compensated and the shear in CY1 in Fig. \ref{fig:BF_Images}(c) and (d) is rectified in Fig. \ref{fig:BF_Images} (g) and (h). The mean GDS and the gCNR metrics validated the proposed FM based DAS and the results are given in Table \ref{tab:compare_metric}, row 2. The FM based DAS is able to correctly localize atleast 70\% of the point targets in all cases except M4 where it is slightly reduced (60\%). However, the improvement is significant when compared to the conventional DAS. The same consistency is also observed in the gCNR values. The results show the capability of the proposed approach in the accurate estimation of delay in DAS beamforming given the ground truth SoS distribution of the medium. The presented study highlights the significance of considering the heterogeneity in SoS during beamforming and is expected to improve the outcomes in US imaging, particularly in tissues with fat depositions. The proposed algorithm is implemented on an Intel(R) Core (TM) i7-8700 CPU (3.20 GHz) workstation and takes an average of 90 s of compute time. The experimental validations of the proposed work, estimation of approximate SoS map (leveraging from the extensive literature) and acceleration of the fast marching algorithm will be considered as a future work. 

\addtolength{\textheight}{-12cm}   

\section*{ACKNOWLEDGMENT}
The authors would like to acknowledge the computing facilities provided by GE Healthcare Bangalore and the  Center for Computational Imaging, Indian Institute of Technology Palakkad, India.

\bibliographystyle{IEEEtran}
\bibliography{root}

\begin{thebibliography}{1}
\providecommand{\url}[1]{#1}
\csname url@samestyle\endcsname
\providecommand{\newblock}{\relax}
\providecommand{\bibinfo}[2]{#2}
\providecommand{\BIBentrySTDinterwordspacing}{\spaceskip=0pt\relax}
\providecommand{\BIBentryALTinterwordstretchfactor}{4}
\providecommand{\BIBentryALTinterwordspacing}{\spaceskip=\fontdimen2\font plus
\BIBentryALTinterwordstretchfactor\fontdimen3\font minus
  \fontdimen4\font\relax}
\providecommand{\BIBforeignlanguage}[2]{{%
\expandafter\ifx\csname l@#1\endcsname\relax
\typeout{** WARNING: IEEEtran.bst: No hyphenation pattern has been}%
\typeout{** loaded for the language `#1'. Using the pattern for}%
\typeout{** the default language instead.}%
\else
\language=\csname l@#1\endcsname
\fi
#2}}
\providecommand{\BIBdecl}{\relax}
\BIBdecl

\bibitem{Szabo2004DiagnosticEdition}
T.~L. Szabo, ``{Diagnostic Ultrasound Imaging: Inside Out: Second Edition},''
  \emph{Diagnostic Ultrasound Imaging: Inside Out: Second Edition}, pp. 1--549,
  2004.

\bibitem{Wang2013TranscranialStudy}
T.~Wang and Y.~Jing, ``{Transcranial ultrasound imaging with speed of
  sound-based phase correction: a numerical study},'' \emph{Physics in Medicine
  {\&} Biology}, vol.~58, no.~19, p. 6663, 9 2013.

\bibitem{Ali2022DistributedEstimates}
R.~Ali, T.~Brevett, D.~Hyun, L.~L. Brickson, and J.~J. Dahl, ``{Distributed
  Aberration Correction Techniques Based on Tomographic Sound Speed
  Estimates},'' \emph{IEEE Transactions on Ultrasonics, Ferroelectrics, and
  Frequency Control}, vol.~69, no.~5, pp. 1714--1726, 5 2022.

\bibitem{HoelRindal2018AModel}
O.~M. Hoel~Rindal, A.~Rodriguez-Molares, and A.~Austeng, ``{A Simple, Artifact
  - Free, Virtual Source Model},'' \emph{IEEE International Ultrasonics
  Symposium, IUS}, vol. 2018-October, 12 2018.

\bibitem{Sethian20123-DMethod}
J.~A. Sethian and A.~M. Popovici, ``{3-D traveltime computation using the fast
  marching method},'' \emph{Geophysics}, vol.~64, no.~2, pp. 516--523, 2 2012.

\bibitem{Treeby2010K-Wave:Fields}
B.~E. Treeby and B.~T. Cox, ``{k-Wave: MATLAB toolbox for the simulation and
  reconstruction of photoacoustic wave fields},'' \emph{J. Biomed. Opt.},
  vol.~15, no.~2, p. 021314, 3 2010.

\bibitem{Liebgott2016Plane-WaveUltrasound}
H.~Liebgott, A.~Rodriguez-Molares, F.~Cervenansky, J.~A. Jensen, and
  O.~Bernard, ``{Plane-Wave Imaging Challenge in Medical Ultrasound},''
  \emph{IEEE International Ultrasonics Symposium, IUS}, vol. 2016-November, 11
  2016.

\bibitem{Rodriguez-Molares2020TheDetectability}
A.~Rodriguez-Molares, O.~M.~H. Rindal, J.~D'Hooge, S.~E. Masoy, A.~Austeng,
  M.~A. Lediju~Bell, and H.~Torp, ``{The Generalized Contrast-to-Noise Ratio: A
  Formal Definition for Lesion Detectability},'' \emph{IEEE Transactions on
  Ultrasonics, Ferroelectrics, and Frequency Control}, vol.~67, no.~4, pp.
  745--759, 4 2020.

\bibitem{AccurateCentral}
\BIBentryALTinterwordspacing
``{Accurate Fast Marching - File Exchange - MATLAB Central}.'' [Online].
  Available:
  \url{https://uk.mathworks.com/matlabcentral/fileexchange/24531-accurate-fast-marching}
\BIBentrySTDinterwordspacing

\end{thebibliography}

\end{document}